# Anomalous Momentum States, Non-Specular Reflections, and Negative Refraction of Phase-Locked, Second Harmonic Pulses

Vito Roppo[1], Marco Centini[2], Domenico de Ceglia[3], Maria Antonietta Vicenti[3], Joseph W. Haus[4], Neset Akozbek[5], Mark J. Bloemer[5], Michael Scalora[5*]

*1  Departament de Fisica i Enginyeria Nuclear, Universitat Politecnica de Catalunya C/Colom 11, 08222 Terrassa Spain*

*2  Dipartimento di Energetica, University of Rome La Sapienza, Via Scarpa 16, Rome Italy*

*3 Dipartimento di Elettrotecnica ed Elettronica, Politecnico di Bari, Via Orabona 4, 70125 Bari, Italy*

*4  Electro-Optics Program, University of Dayton, Dayton, OH 45469-0245, USA*

*5  Charles M. Bowden Research Facility, RDECOM, US Army Aviation and Missile Command, Redstone Arsenal, AL 35803*

*\* Corresponding Author*

## Abstract

We simulate and discuss novel spatio-temporal propagation effects that relate specifically to pulsed, phase-mismatched second harmonic generation in a negative index material having finite length.  Using a generic Drude model for the dielectric permittivity and magnetic permeability, the fundamental and second harmonic frequencies are tuned so that the respective indices of refraction are negative for the pump and positive for the second harmonic signal.  A phase-locking mechanism causes part of the second harmonic signal generated at the entry surface to become trapped and dragged along by the pump and to refract negatively, even though the index of refraction at the second harmonic frequency is positive. These circumstances culminate in the creation of an anomalous state consisting of a forward-moving second harmonic wave packet that has negative wave vector and momentum density, which in turn leads to non-specular reflections at intervening material interfaces. The forward-generated second harmonic signal trapped under the pump pulse propagates forward, but has all the attributes of a reflected pulse, similar to its twin counterpart generated at the surface and freely propagating backward away from the interface. This describes a new state of negative refraction, associated with nonlinear frequency conversion and parametric processes, whereby a beam generated at the interface can refract negatively even though the index of refraction at that wavelength is positive.

The primary reason to investigate second harmonic generation (SHG) [1] has been the achievement of efficient frequency doubling, with emphasis on phase-matched interactions. In bulk media, phase matching (PM) is a momentum-conserving condition that does not generally occur naturally. As a result, most efforts have been directed toward the achievement of PM by introducing geometrical modifications aimed at circumventing natural material dispersion and creating effective phase matching conditions [2-24]. The systematic study of nonlinear propagation phenomena at or very near PM conditions has resulted in relatively few studies of SHG and related propagation phenomena in a phase-mismatched (PMM) environment, although from time to time workers have addressed the issue [4,25-31]. The present investigation is an extension of previous work [32], where the presence of a second harmonic component in the output signal was framed in the context of a generic phase-locking mechanism that appears to characterize parametric processes in ordinary as well as negative index materials, or NIMs, including third order processes [33].

Typically, in ordinary, positive index materials (PIMs) the generated second harmonic signal consists of two pulses. One of the pulses is trapped and dragged along by the pump pulse, and propagates under the pump envelope at the pump's group velocity. The second pulse decouples from the pump and propagates with the characteristic group velocity of the second-harmonic frequency, and obeys material dispersion. Although the qualitative aspects of the interaction in both PIMs and NIMs appear to evolve in somewhat parallel tracks, in NIMs the effects of phase-locking manifest themselves in much more dramatic fashion. For example, similar to what occurs in PIMs, in NIMs the pump pulse also captures part of the SH signal, and both co-propagate at the same group velocity, *regardless of material dispersion*. Mirroring what also occurs in PIMs, the relative amplitudes of the electric and magnetic fields change to accommodate the fact that while material dispersion does not allow it, the effective impedance that the pump and the SH fields experience must be the same as a result of phase locking. However, the pump is refracting negatively, and as a result the most consequential adjustment the SH field must make relates to the fact that it is tuned to a region of positive index: its captivity relegates the SH signal to do as the pump does, i.e.

it must refract as if its index of refraction were negative. Moreover, the second harmonic E and H fields no longer overlap [32] even though material dispersion is such that $\varepsilon=\mu$ (this equality renders Maxwell's equations symmetric, E and H fields overlap, and the fields indistinguishable).

Under phase-mismatched conditions, SHG takes place as the pump pulse transits across material entry or exit surfaces in both PIMs and NIMs. The process is characterized by the generation of a backward pulse, and forward-propagating phase-locked and normal pulses. The analysis in reference [32] strongly hinted at the fact that in NIMs reflected and phase-locked pulses may in reality be the same pulse, part of which is emitted backward at the interface and the rest is appropriated and captured by the pump. Here we provide further evidence that these pulses are indeed strongly correlated by numerically showing that the momentum density of the phase-locked pulse is negative, even though the pulse propagates along the positive direction, thus appearing to have the properties of its twin reflected pulse.

In order to understand how these effects come about it helps to study pulsed second harmonic generation under phase matched conditions [34] and compare with the present case. If the interaction were to occur at or very near phase matching, a broad, *backward-propagating* reflected SH pulse is continuously generated under the forward-moving pump pulse deep inside the medium. Therefore, the SH signal is generated under the pump with negative momentum and rushes backward toward the entry surface. Outside of phase matching conditions, the pump generates a reflected SH signal part of which is able to immediately leave the medium, while part of it is generated inside the medium but is not able to escape as it becomes trapped by the pump.

The phase-locking phenomenon just described is exemplified in Figs.1-2 for a PIM, and in Figs.3-4 for a NIM with a normally incident pump pulse. In the figures we show the pulses in physical space and their respective k-space spectra after the pump pulse has transited the entry interface. Remarkably, the introduction of even a vanishingly small discontinuity in a NIM material, for example, turning $\chi^{(2)}$ off somewhere along the longitudinal coordinate, is enough to dislodge the trapped SH pulse from under the pump and send it propagating backward [32]. Figs.5-7 represent before-during-after snapshots of the dynamics that ensues as the pair of phase locked pump and second harmonic pulses crosses from a non-zero $\chi^{(2)}$ region into a region where $\chi^{(2)}=0$, assuming normal incidence. Similar dynamics occurs for a material discontinuity of arbitrary index jump. The sequence gives a clear indication that the phase locking phenomenon is born out of the nonlinear material, which links the harmonics. A connection between the trapped pulse and the backward-generated second harmonic pulses may be established by performing an analysis of the signal in the neighborhood of the phase matching condition, where the correlation between the two pulses becomes more evident. This dynamics has no apparent counterpart in PIMs and is discussed in details in reference [32].

In reference [32] propagation phenomena were studied by integrating Maxwell's equations using one spatial dimension and time, with particular attention given to the dynamics of short femtosecond pulses crossing a small number of positive/negative index interfaces. In the present simulations we consider the addition of a transverse coordinate in order to study the dynamics of incident and generated pulses when the pump pulse impacts the interface(s) at oblique incidence. Consideration of a transverse coordinate facilitates the simultaneous study of diffraction and refraction and, as we will see, it adds a remarkable phenomenon to the character of the interaction. Our results may be summarized as follows: An analysis of the scattered fields reveals that the momentum density of the forward-moving, phase-locked SH pulse is negative. A spectral decomposition also reveals that the trapped SH signal is phase-locked to the pump pulse, which refracts negatively. Taken together, these circumstance lead to an unusual negative refraction state and just as atypical non-specular reflections of the SH signal.

The permittivity and the magnetic permeability of the NIM are described by a standard, lossy Drude model [5]. In Fig.8 we show several snapshots of a pump pulse incident from vacuum on a NIM substrate so that $\varepsilon(\omega)= \mu(\omega) \sim -1.04+i2 \times 10^{-4}$, and $\varepsilon(2\omega)= \mu(2\omega) \sim 0.49+i2 \times 10^{-5}$, for $\omega=0.7$. We assume that both the pump and SH fields are TE polarized, and $\chi^{(2)} \sim 1$pm/V in the region delimited by the left-most, light blue and yellow lines, and zero otherwise. The pump remains undepleted, and so its propagation remains linear throughout the NIM volume. In the sequence, the pulse is seen to refract negatively, as expected.

In Fig.9 we show several corresponding snapshots of the scattered SH fields. For example, in snapshot number 3 at least three SH components are visible: (i) a reflected pulse; (ii) a normal pulse, which refracts downward according to Snell's law. This pulse descends down the page because the index of refraction at the second harmonic frequency is less then unity; and (iii) a phase-locked pulse that follows the same trajectory as the pump, i.e. it refracts negatively even though the index of refraction is positive. As the interaction proceeds forward, the pump crosses the yellow line and spills into a NIM region where $\chi^{(2)}=0$. This event turns off the coupling between the pump and the trapped SH pulse and, similarly to what occurs in one dimension [32], the SH pulse is released by the pump. Later snapshots clearly show that the phase-locked, SH signal does not reflect specularly at the interface. Rather, it is retro-reflected, following its twin pulse generated at the first vacuum/NIM interface.

The situation depicted in Fig.9 is extraordinary because it shows that a pulse tuned to a region of positive index acts as if its index of refraction were negative, thus pointing to a new way to arrive at negative refraction. In order to understand the retro-reflection, in Fig.10 we depict the longitudinal momentum density (Poynting vector) $S_z(y,z)$ that corresponds to snapshot number 3 of Fig.9. The figure shows that the calculated momentum density is negative not only for the reflected pulse, as expected, but also for the forward-propagating phase-locked pulse. In Fig.11 we show the transverse momentum density $S_y(y,z)$, and there again we see that the phase-locked, SH pulse has no positive components. Therefore, since the transverse momentum component must be conserved due to the lack of transverse boundaries, and since both components are negative, the pulse has no choice but to return whence it came.

The k-space spectral decomposition of Fig.4 implies that the carrier wave vector of the phase-locked SH pulse is negative as a result of phase-locking to the pump. This statement implies that the pump and the SH signal propagate with the same effective index. This situation is clearly anomalous, and causes the E and H fields of the phase-locked pulse to acquire different amplitudes, as shown in Fig.3, for example. Therefore, the characteristics of the phase-locked pulse are identical to the properties of the reflected pulse, which has negative k-vector and negative momentum density. In fact, what occurs at the surface may be characterized as follows: as the pump pulse enters the medium it generates a backward propagating SH pulse, part of which immediately finds its way out, while a portion is generated under the pump just inside the entry surface with negative momentum. Once the pump advances further into the medium, it schleps and retains that portion of the SH signal until a second scattering event occurs that frees the trapped pulse. The situation described is remarkable because it points to a situation where the strong relationship that rapidly ensues between the fundamental and phase-locked pulse occurs *regardless* of material dispersion and the kind of material under consideration. In fact we see the effect develop for second and third order processes and for different harmonics [32, 33]. Moreover, there are indications that phase-locking and trapping also occur for seeded dynamics, once again regardless of the kind of material at hand, or the degree of phase mismatch. Therefore, the phase-locking mechanism appears to be a broad-based, fundamental effect that characterizes parametric processes, perhaps even including down-conversion.

In summary, we have discussed a novel way to obtain negative refraction, by bringing to bear nonlinear parametric processes in negative index materials. The introduction of a transverse coordinate makes possible the simultaneous study of diffraction and refraction across a PIM/NIM interface, and shows that a second harmonic signal tuned to a region of positive index can refract negatively if it becomes trapped by a negatively refracting pump pulse. As the pump crosses the interface a set of twin pulses is generated, one that is immediately reflected, the other becomes trapped by the pump and propagates deep inside the medium. However, we have shown that the properties of the forward-moving SH pulse are identical to the characteristics of its twin pulse, as both display negative wave vector and momentum density. Therefore, once the trapped pulse reaches an interface, the interaction ceases, the phase-locked pulse frees itself of the pump's grip, and thanks to conservation of transverse momentum it is retro-reflected. Finally, we note that the results we have presented clearly reveal that many subtleties relating to ultrashort pulse propagation and the phase locking mechanism are still not well understood. Nevertheless, the results here also provide new vistas that may lead to the development of novel devices. An examination of Fig.9 brings to mind a phase conjugate mirror, for example, although we believe that many more pleasant surprises await to be revealed.

Acknowledgement: we thank the US Army-European Research Office and the US National Research Council for Partial Financial Support

**References**

[1] P. A. Franken, A. E. Hill, C. W. Peters, and G. Weinreich, "Generation of Optical Harmonics", Phys. Rev. Lett. **7**, 118 (1961).

[2] J. A. Giordmaine, "Mixing of Light Beams in Crystals", Phys. Rev. Lett. **8**, 19 (1962).

[3] P. D. Maker, R.W. Terhune, M. Nisenoff, and C. M. Savage, "Effects of Dispersion and Focusing on the Production of Optical Harmonics", Phys. Rev. Lett. **8**, 21 (1962).

[4] J. A. Armstrong, N. Bloembergen, J. Ducuing, and P. S. Pershan, "Interactions between Light Waves in a Nonlinear Dielectric", Phys. Rev. **127**, 1918 (1962); N.

Bloembergen, and P. S. Pershan, "Light Waves at the Boundary of Nonlinear Media", Phys. Rev. **128**, 606 (1962).

[5] K. Beckwitt, Y.F. Chen, F.W. Wise and B. A. Malomed, "Temporal solitons in quadratic nonlinear media with opposite group-velocity dispersions at the fundamental and second harmonics", Phys. Rev. E **68**, 057601 (2003).

[6] M. M. Fejer, G. A. Magel, D. H. Jundt, and R. L. Byer, "Quasi-phase-matched second harmonic generation: tuning and tolerances," IEEE J. Quant. Electron. **28**, 2631 (1992).

[7] U. Sapaev and D. Reid, "General second-harmonic pulse shaping in grating-engineered quasi-phase-matched nonlinear crystals", Opt. Express **13**, 3264 (2005).

[8] G. D. Boyd and C. K. N. Patel, "Enhancement of optical second-harmonic generation (SHG) by reflection phase matching in ZnS and GaAs", App. Phys. Lett. **8**, 12 (1966).

[9] E. V. Petrov and B. I. Mantsyzov, "Modified phase matching conditions for second harmonic generation in a finite one-dimensional photonic crystal near the bragg condition: Weak and strong diffraction", Journal of Experimental and Theoretical Physics **101,** 3 (2005).

[10] M. G. Weber, "Enhanced second-harmonic generation by surface polaritons on thin metal films", Phys. Rev. B **33**, 6775 (1986).

[11] E. Sidick, A. Knoesen and A. Dienes, "Ultra-short pulse second harmonic generation in quasi-phase matched structures", Pure Appl. Opt. **5**, 709 (1996).

[12] D. T. Reid, "Engineered quasi-phase-matching for second-garmonic generation", J. Opt. A Pure Appl. Opt. **5**, S97 (2003).

[13] A. Steinbach, M. Rauner, F.C. Cruz and J.C. Bergquist, "CW second harmonic generation with elliptical Gaussian beams", Opt. Comm. **123**, 207 (1996).

[14] G. D'Aguanno, N. Mattiucci, M. J. Bloemer and M. Scalora, "Accessing quadratic nonlinearities of metals through metallo-dielectric photonic-band-gap structures", Phys. Rev. E **74**, 036605 (2006).

[15] K. Gallo, P. Baldi, M. De Micheli, D. B. Ostrowsky and G.Assanto, "Cascading phase shift and multivalued response in counterpropagating frequency-nondegenerate parametric amplifiers", Opt. Lett. **25**, 13 (2000).

[16] C. Conti, G. Assanto and S. Trillo, "Cavityless oscillation through backward quasi-phase-matched second-harmonic generation", Opt. Lett. **24**, 16 (1999).

[17] N. Fujioka, S. Ashihara, H. Ono, T. Shimura and K. Kuroda, "Group-velocity-mismatch compensation in cascaded third-harmonic generation with two dimensional quasi-phase-matching gratings", Opt. Lett. **31**, 18 (2006).

[18] M. Scalora, M. J. Bloemer, A. S. Manka, J. P. Dowling, C. M. Bowden, R. Viswanathan and J. W. Haus, "Pulsed second-harmonic generation in nonlinear, one-dimensional, periodic structures", Phys. Rev. A **56**, 4 (1997).

[19] M. Centini, G. D'Aguanno, M. Scalora, C. Sibilia, M. Bertolotti, M. J. Bloemer and C. M. Bowden, "Simultaneously phase-matched enhanced second and third harmonic generation", Phys. Rev. E **64**, 046606 (2001).

[20] T. Wang, L. Qian, H. Zhu, Z. Hou and F. Li "Beam-quality improved efficient second harmonic generation in a periodic quasi-phase-matching gratings", Opt. Comm. **188**, 213 (2001).

[21] V. Krylov, A. Rebane, A. G. Kalintsev, H. Schwoerer and U. P. Wild, "Second-harmonic generation of amplified femtosecond Ti:sapphire laser pulses", Opt. Lett. **20**, 2 (1995).

[22] Y. Liu and J. C. Diels, "Group-Velocity Matched Femtosecond Parametric Oscillation by Noncollinear Quasi-Phase Matching", IEEE J. Quant. Electron. **42**, 8 (2006).

[23] F. Xu, J. L. He, J. Liao, Q. Wang, Q. Xu, N. H. Shen, H. T. Wang and N. B. Ming, "Simultaneous high-efficiency and equal-level second- and third-harmonic generation achieved by controllable linear gratings in a quasiperiodic optical superlattice", Phys. Rev. A **68**, 053803 (2003).

[24] S. Orlov, A. Yariv and M. Segev, "Nonlinear self-phase matching of optical second harmonic generation in lithium niobate", Appl. Phys. Lett. **68**, 12 (1996).

[25] S.L. Shapiro, "Second-harmonic generation in $L_iN_bO_z$ by picosecond pulses", Appl. Phys. Lett. **13**, 19 (1968).

[26] W.H. Glenn, "Second-Harmonic Generation bv Picosecond Optical Pulses ", IEEE J. Quant. Electron. **QE-5**, 6 (1969).

[27] M. M. Fejer, "Nonlinear optical frequency conversion", Phys. Today **47**, 25 (1994).

[28] J. T. Manassah and O. R. Cockings, "Induced phase modulation of a generated second-harmonic Signal", Opt. Lett. **12,** 12 (1987).

[29] L. D. Noordam, H. J. Bakker, M. P. de Boer, and H. B. van Linden van den Heuvell, "Second-harmonic generation of femtosecond pulses: observation of phase-mismatch effects", Opt. Lett. **15**, 24 (1990).

[30] R. Maleck Rassoul, A. Ivanov, E. Freysz, A. Ducasse and F. Hache, "Second-harmonic generation under phase-velocity and group-velocity mismatch: influence of cascading self-phase and cross-phase modulation", Opt. Lett. **22**, 268 (1997).

[31] W. Su, L. Qian, H. Luo, X. Fu, H. Zhu, T. Wang, K. Beckwitt, Y. Chen and F. Wise, "Induced group-velocity dispersion in phase-mismatched second-harmonic generation", J. Opt. Soc. Am. B **23**, 51 (2006).

[32] V.Roppo, M. Centini, C. Sibilia, M. Bertolotti, D.de Ceglia, M. Scalora, N.Akozbek, M. J. Bloemer, J. W. Haus, O. G. Kosareva and V. P. Kandidov, "Role of phase matching in pulsed second-harmonic generation: Walk-off and phase-locked twin pulses in negative-index media", Phys. Rev. A **76,** 033829 (2007)

[33] N. Aközbek, A. Iwasaki, A. Becker, M. Scalora, S. L. Chin and C. M. Bowden, "Third-Harmonic generation and Self-Channeling in Air Using High-Power Femtosecond Laser Pulses", Phys. Rev. Lett. **89**, 14 (2002).

[34] M. Scalora, G. D'Aguanno, M. Bloemer, M. Centini, D. de Ceglia, N. Mattiucci and Y. S. Kivshar, "Dynamics of short pulses and phase matched second harmonic generation in negative index materials", Opt. Express **14**, 4746 (2006).

**Figure Captions**

**Fig.1:** A Gaussian pump pulse -scale on left axis- propagates into a positive index material having $n(\tilde{\omega}) = 1.47$ and $n(2\tilde{\omega}) = 1.64$ at the fundamental frequency $\tilde{\omega} = 0.84$. The second harmonic signal - scale on right axis - is characterized by a reflected pulse, and two forward-propagating pulses, one traveling at the normal group velocity, which walks off and lags behind, the other phase-locked, and located under the pump and traveling at the pump's group velocity. Here too, $\chi^{(2)} \sim 1 \text{pm/V}$, and absorption is negligible.

**Fig.2:** K-space power spectrum for the situation depicted Fig.1. For the pump (FF) -left axis scale- we identify: (*i*) the reflected component ($k_\omega = 2\pi\tilde{\omega}$), and (*ii*) the component that propagates inside the medium ($k_\omega n_\omega$). For the second harmonic (SH) - right axis scale - it is possible to recognize (*i*) the reflected component ($k_{2\omega}$), (*ii*) the component propagating inside the medium that walks off, having nominal group velocity given by material dispersion ($k_{2\omega} n_{2\omega}$), and (*iii*) the phase-locked component ($2k_\omega n_\omega$).

**Fig.3:** A Gaussian pump pulse -black curve, left axis scale- is tuned to $\tilde{\omega} = 0.84$, and propagates into a negative index material such that $n(\tilde{\omega}) = -0.41$ and $n(2\tilde{\omega}) = 0.64$. The second harmonic -red curve, left axis scale- has a reflected pulse, a forward-propagating pulse that leads the way, and propagates at the nominal group velocity for the second harmonic frequency, and a phase-locked pulse that propagates at the group velocity of the fundamental pump field. $\chi^{(2)} = 1\,\text{pm/V}$, and absorption is negligible.

**Fig.4:** K-space power spectrum of the situation in Fig.7. For the pump field -left axis scale- we have a forward-propagating component characterized by a negative carrier wave vector ($k_\omega n_\omega$). Pump reflections are negligible. For the second harmonic wavelength -right axis scale- we identify: (*i*) the reflected component ($-k_{2\omega}$), (*ii*) the normal component that walks off ($k_{2\omega} n_{2\omega}$), and (*iii*) the phase-locked component ($2k_\omega n_\omega$).

**Fig.5:** A Gaussian pulse tuned to $\tilde{\omega} = 0.84$ -black curve, left axis scale- propagates into a negative index material having $n(\tilde{\omega}) = -0.41\, n(\tilde{\omega}) = -0.41$ and $n(2\tilde{\omega}) = 0.64$, just before a $\chi^{(2)}$ interface is crossed. The phase-locked SH pulse -red and blue curves, right

axis scale- propagates at the same group velocity as the FF. $\chi^{(2)} = 1\,\text{pm/V}$, and absorption is negligible.

**Fig.6:** The same pulses of Fig.5 during the $\chi^{(2)}$ interface crossing. Note that the FF pulse crosses undisturbed, while the SH pulse is stopped as a second forward-moving SH pulse begins to form to the right of the interface.

**Fig.7:** The same pulses of Fig.5 after the $\chi^{(2)}$ interface crossing. The phase-locked pulse is now freely propagating backward with the group velocity dictated by material dispersion, and a forward-propagating SH pulse is also clearly visible.

**Fig.8:** A pump pulse is incident at oblique incidence on a NIM, which is divided into linear and nonlinear portions. SHG remains in the undepleted regime, and so the pump propagates feeling no influence from the SH signal.

**Fig.9:** Generated second harmonic pulses. A first pulse is generated backward, while two pulses are seen to depart from the entry surface, one downward, which refracts according to material dispersion and Snell's law, one upward, phase-locked and trapped by the pump pulse. We note that the index of refraction at the SH frequency is positive. Nevertheless, the signal refracts negatively, following the pump pulse. Once the pulse reached the interface that separates a linear from a nonlinear NIM, the SH pulse is freed from the pulse, and is retro-reflected in the direction whence it came.

**Fig.10:** Longitudinal momentum density that corresponds to snapshot number 3 of both Figs.9 and 10. It is evident that the momentum density of the phase-locked, forward propagating pulse is negative, just as the momentum density of the reflected pulse is negative, thus providing further proof that the pulses are indeed twin. The pump pulse

rips apart the pulse generated at the interface, and freezes the portion that it captures and drags along, to release it only at the next interface crossing.

**Fig.11:** Transverse momentum density that corresponds to Fig.10. Here to the momentum components of the phase locked pulse are negative. The lack of any positive momentum components in the phase locked pulse renders specular reflections impossible under the circumstances.

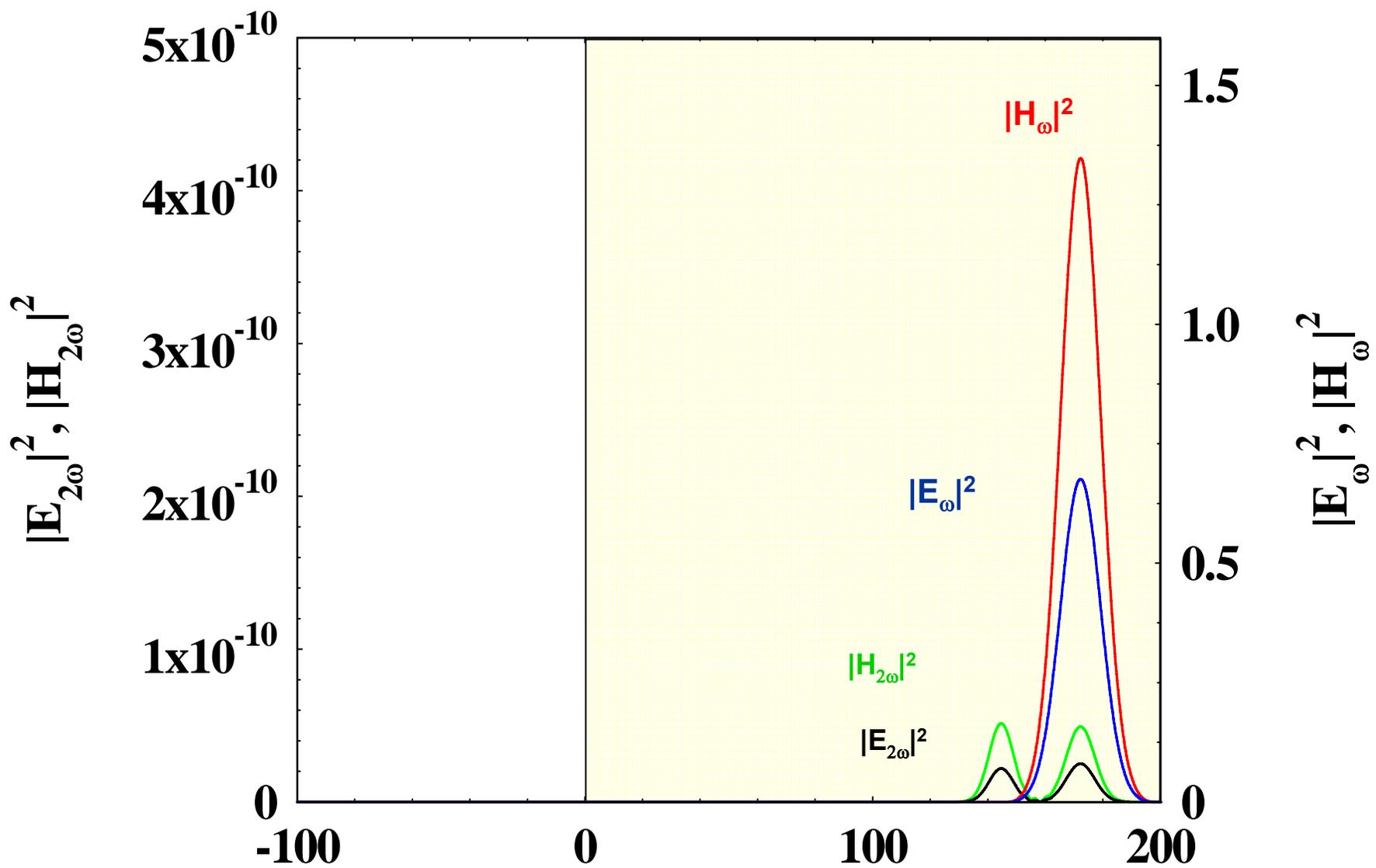

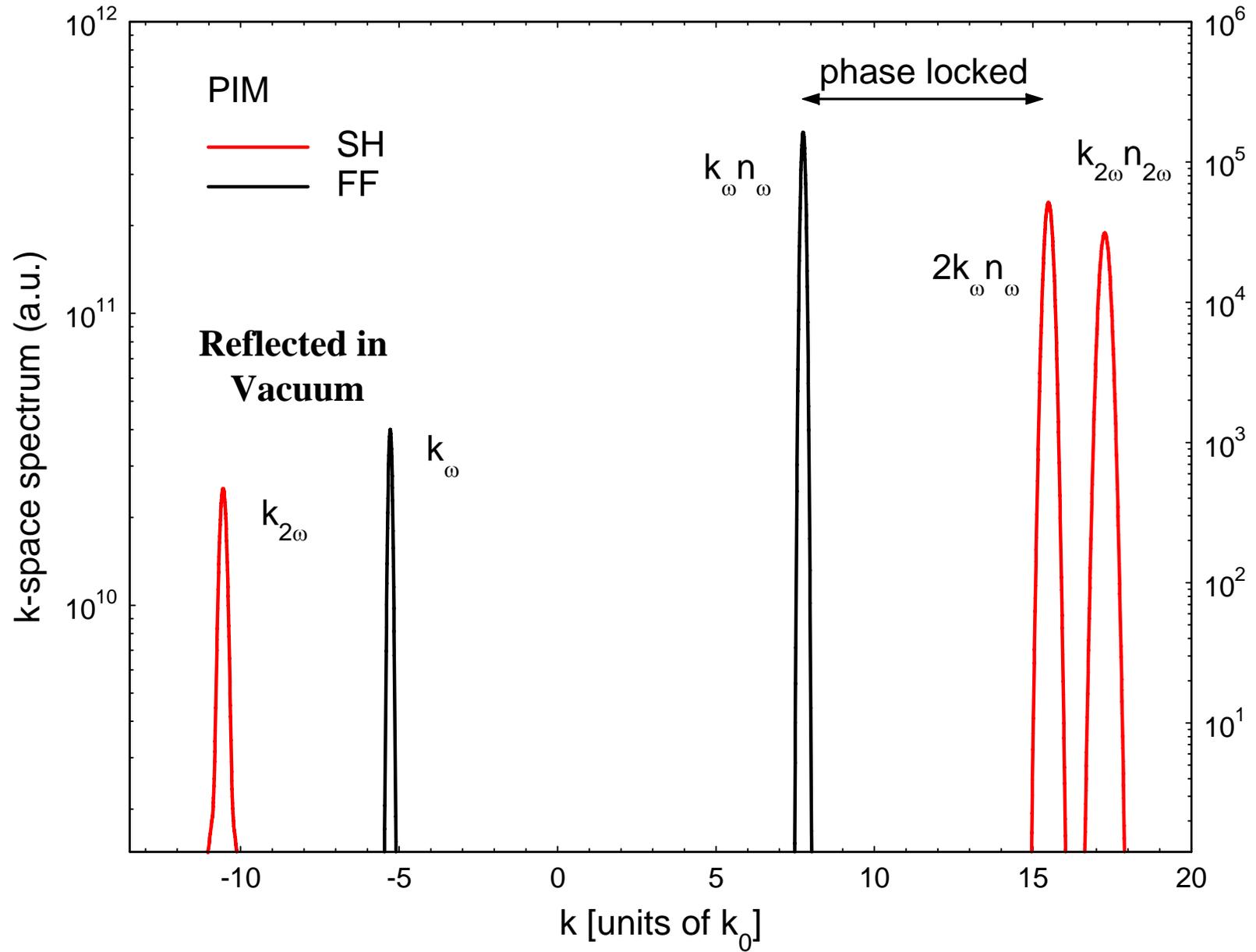
2

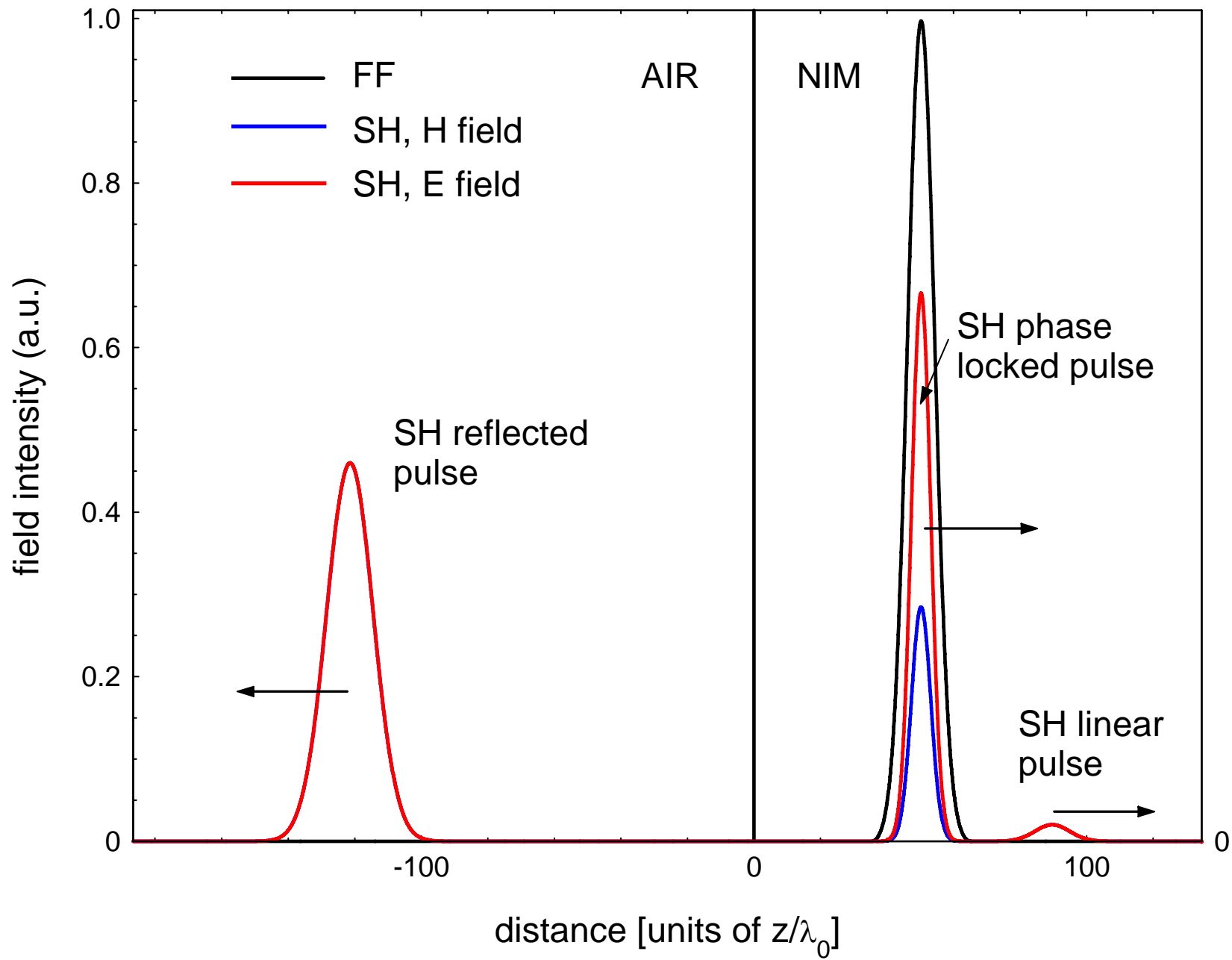

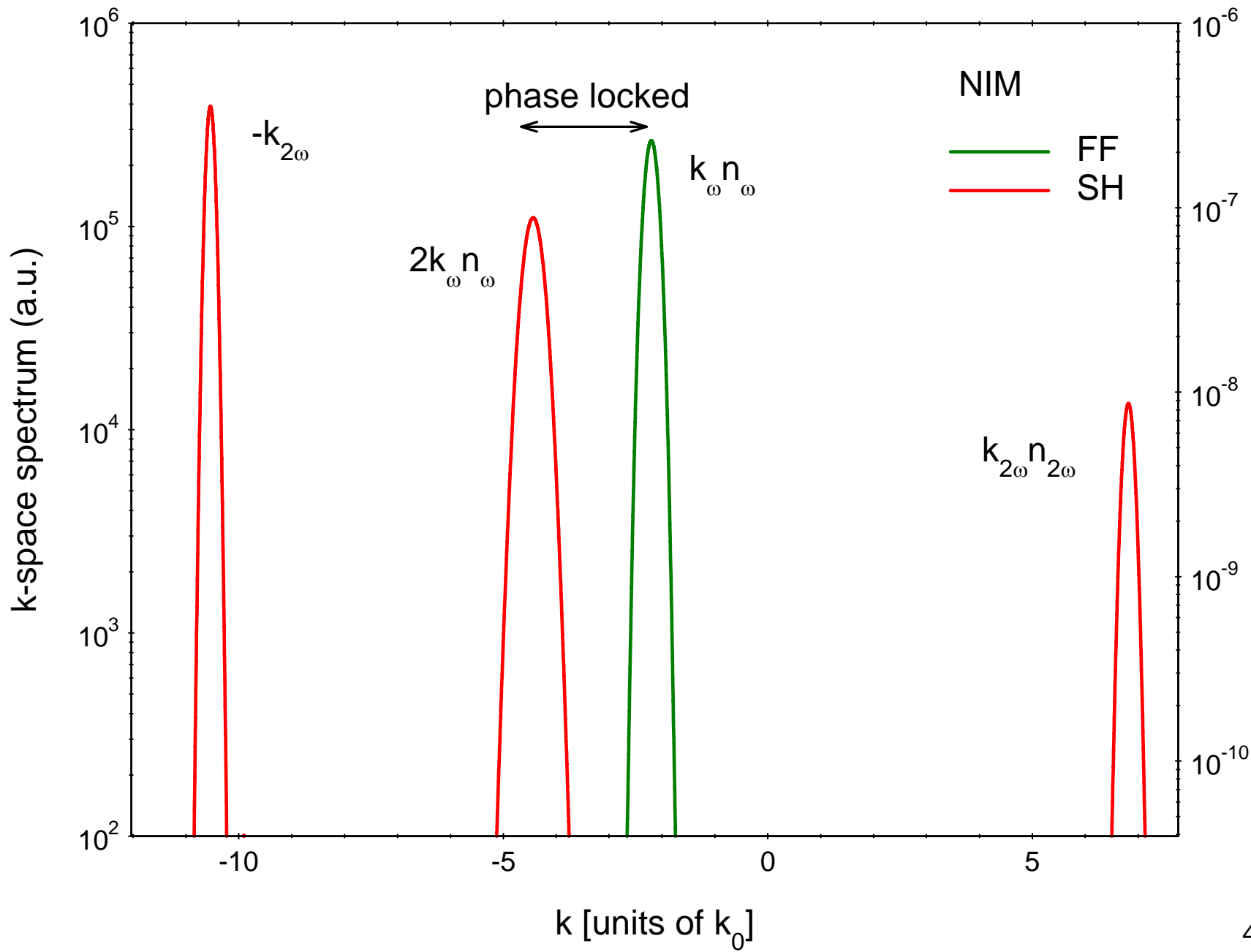

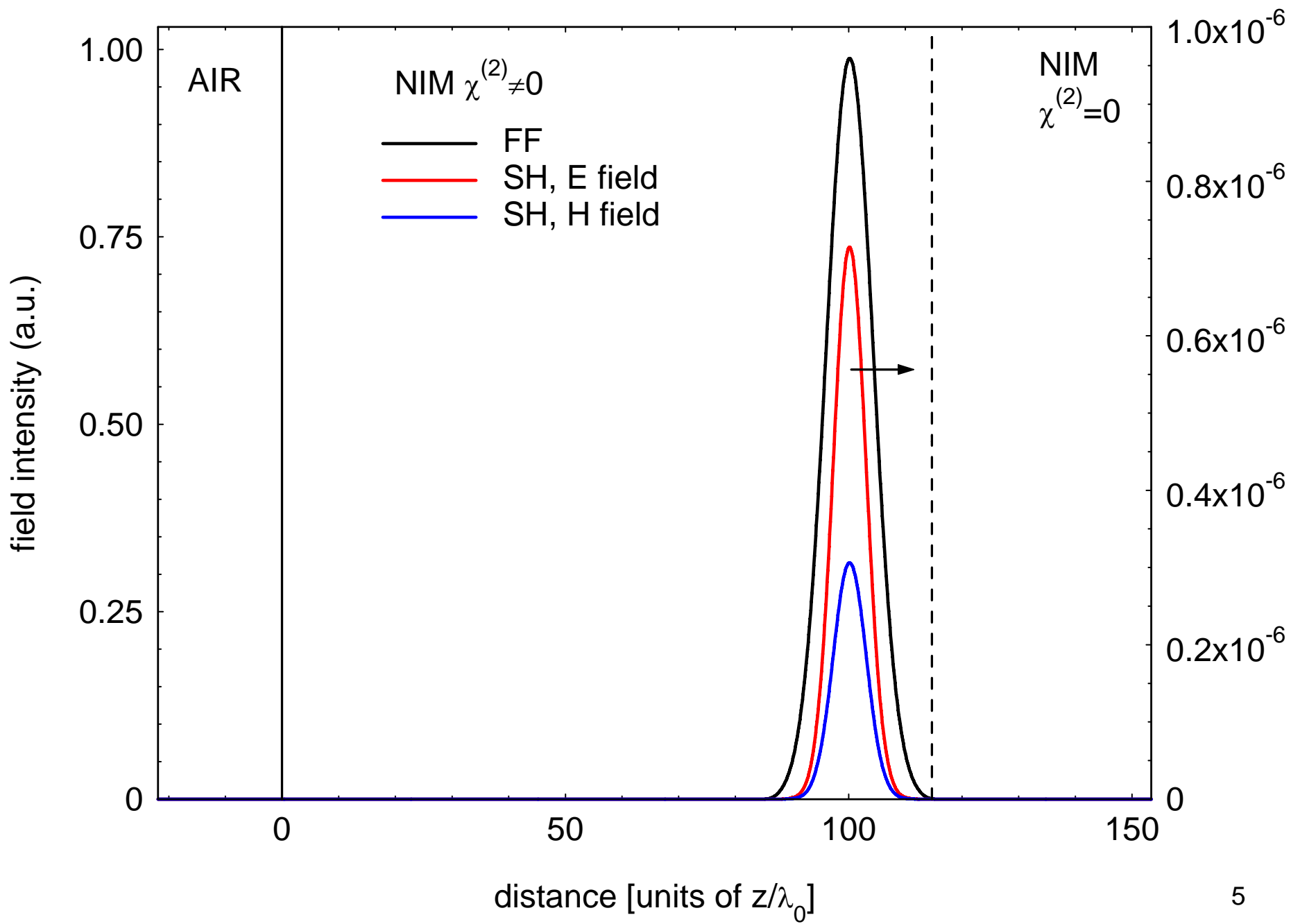

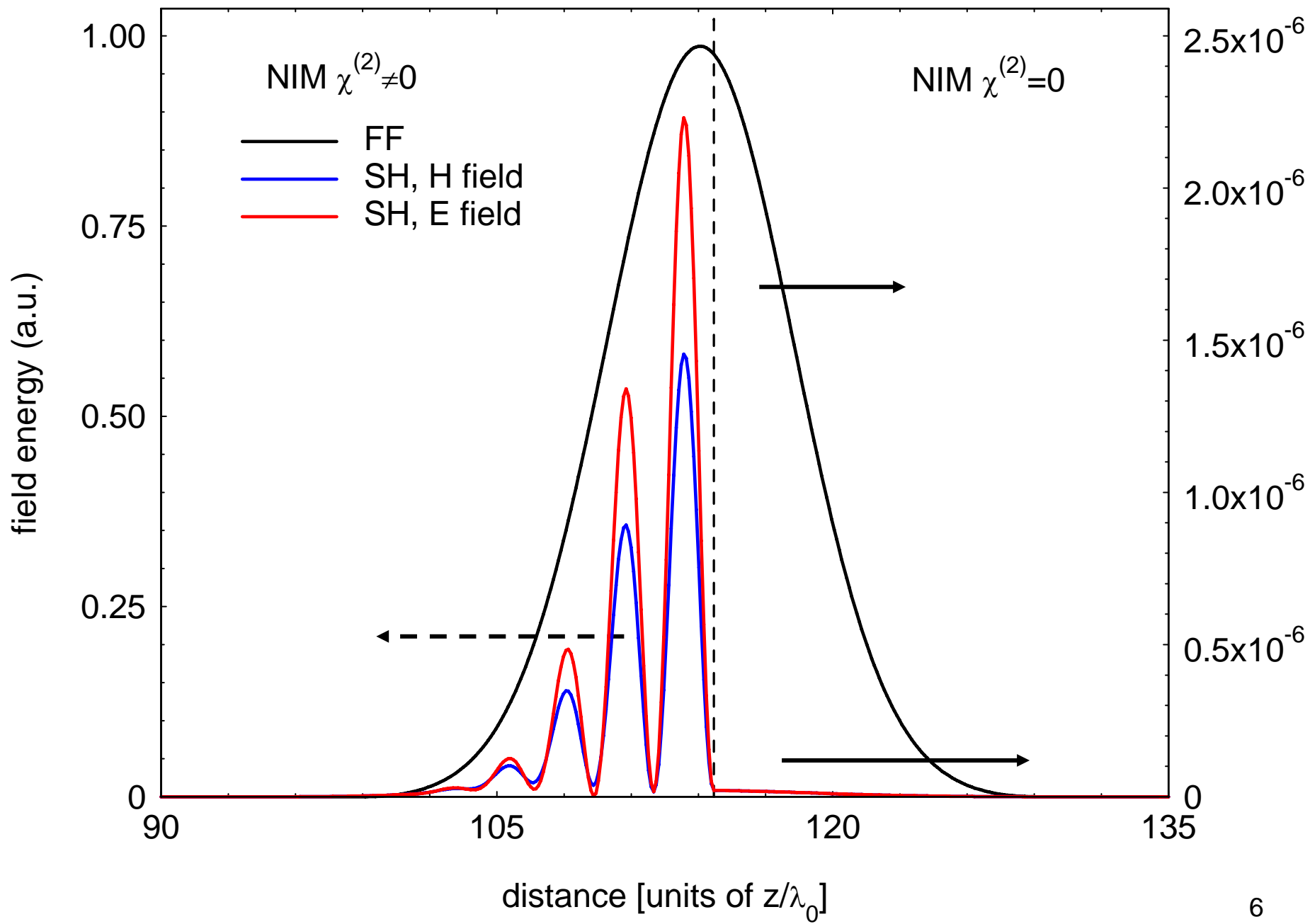

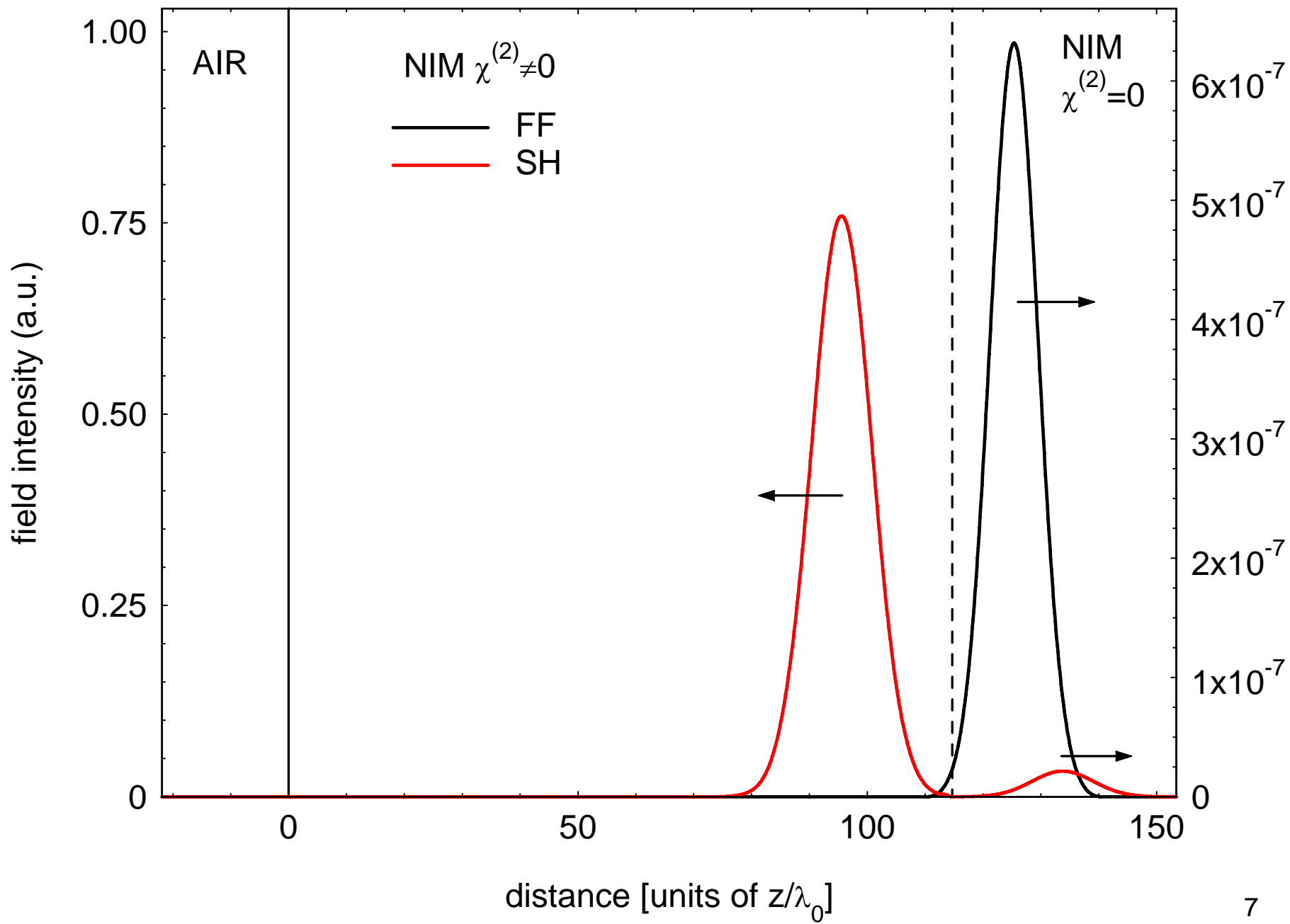

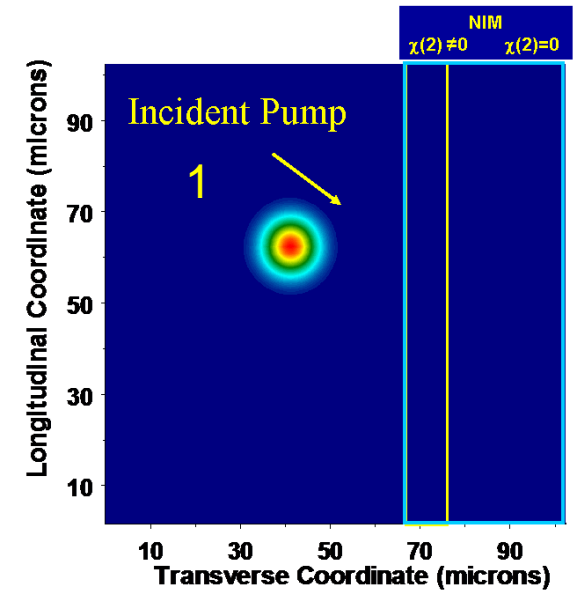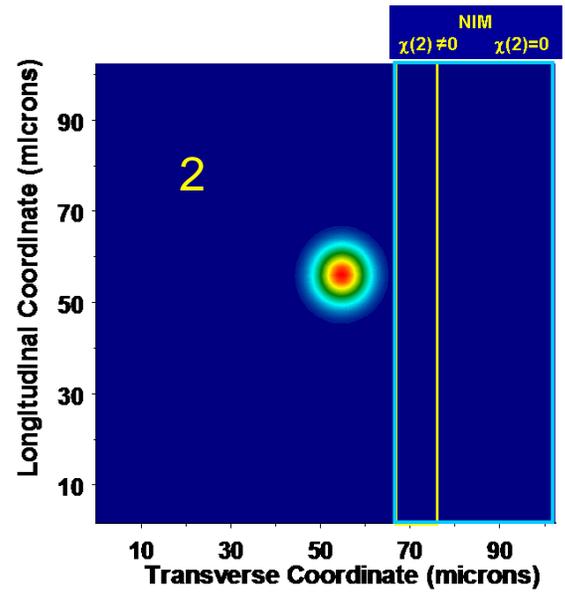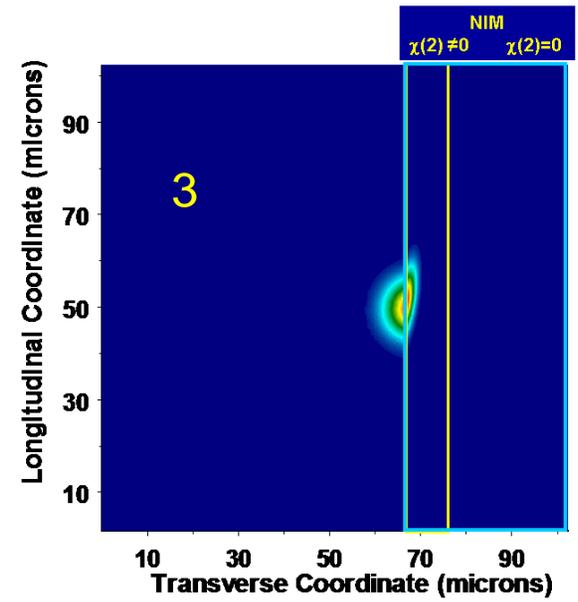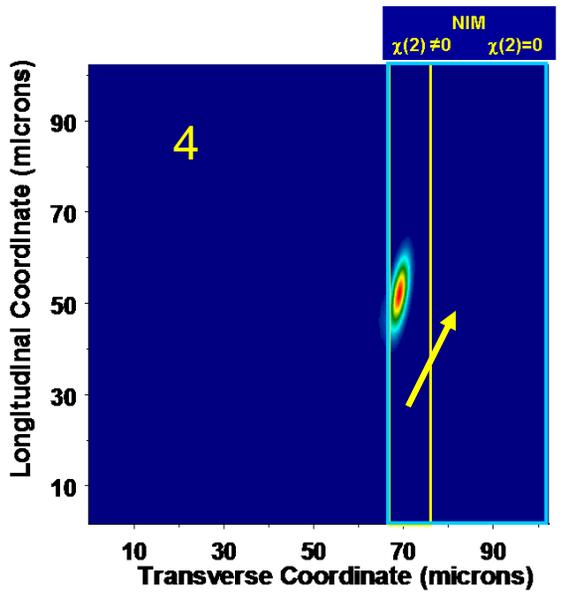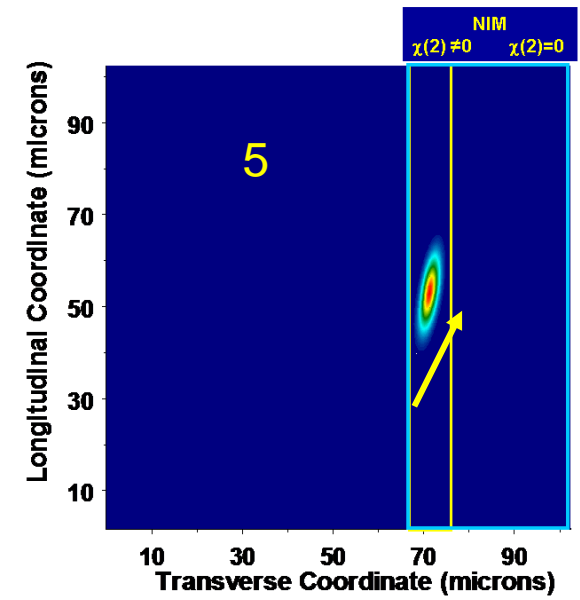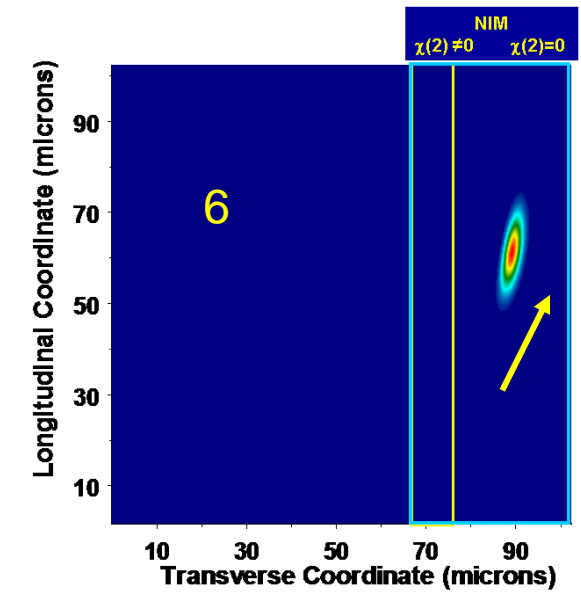

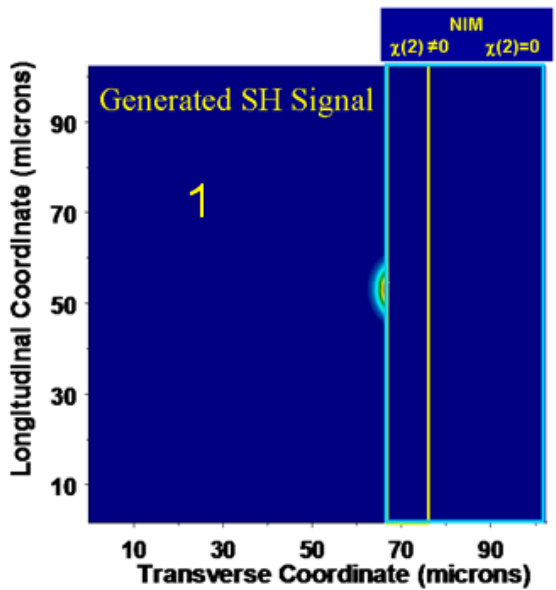
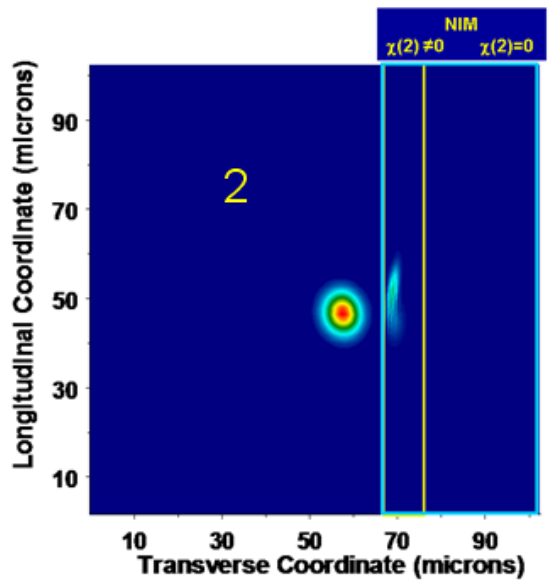
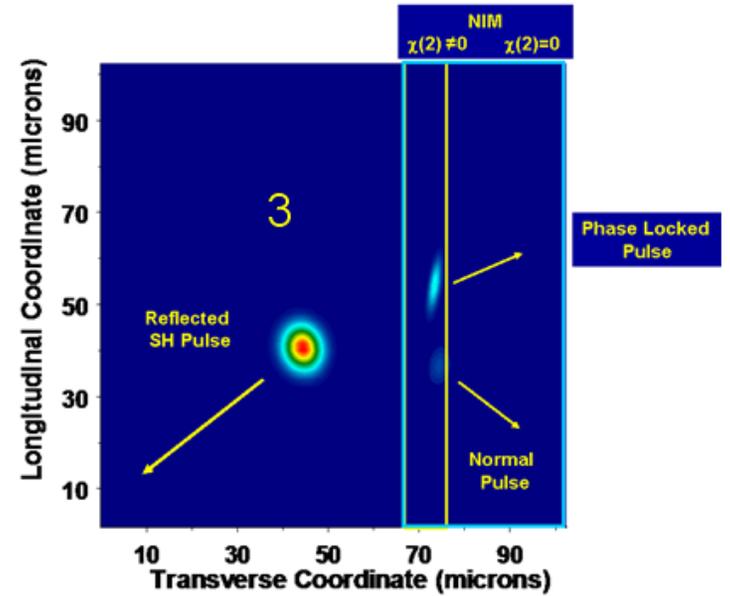
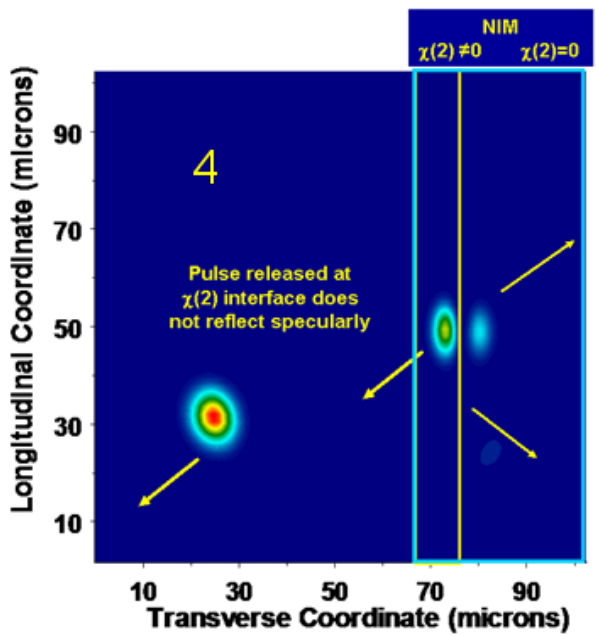
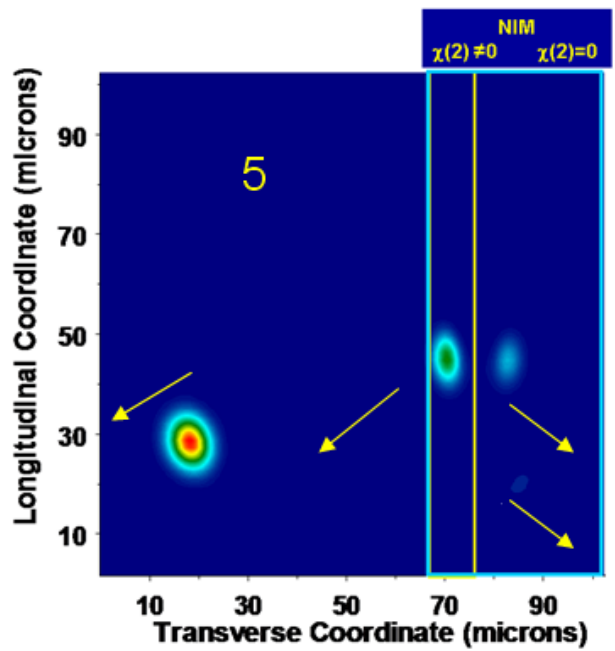
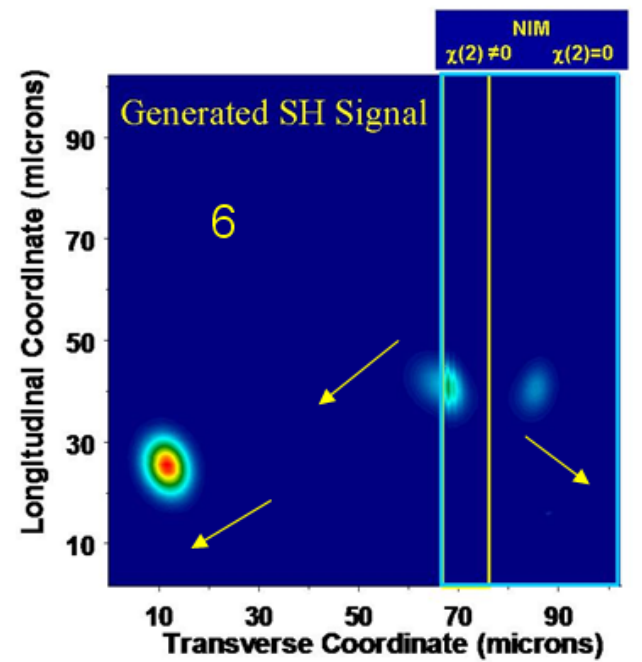

# Longitudinal Poynting Vector  $S_z(2\omega)$

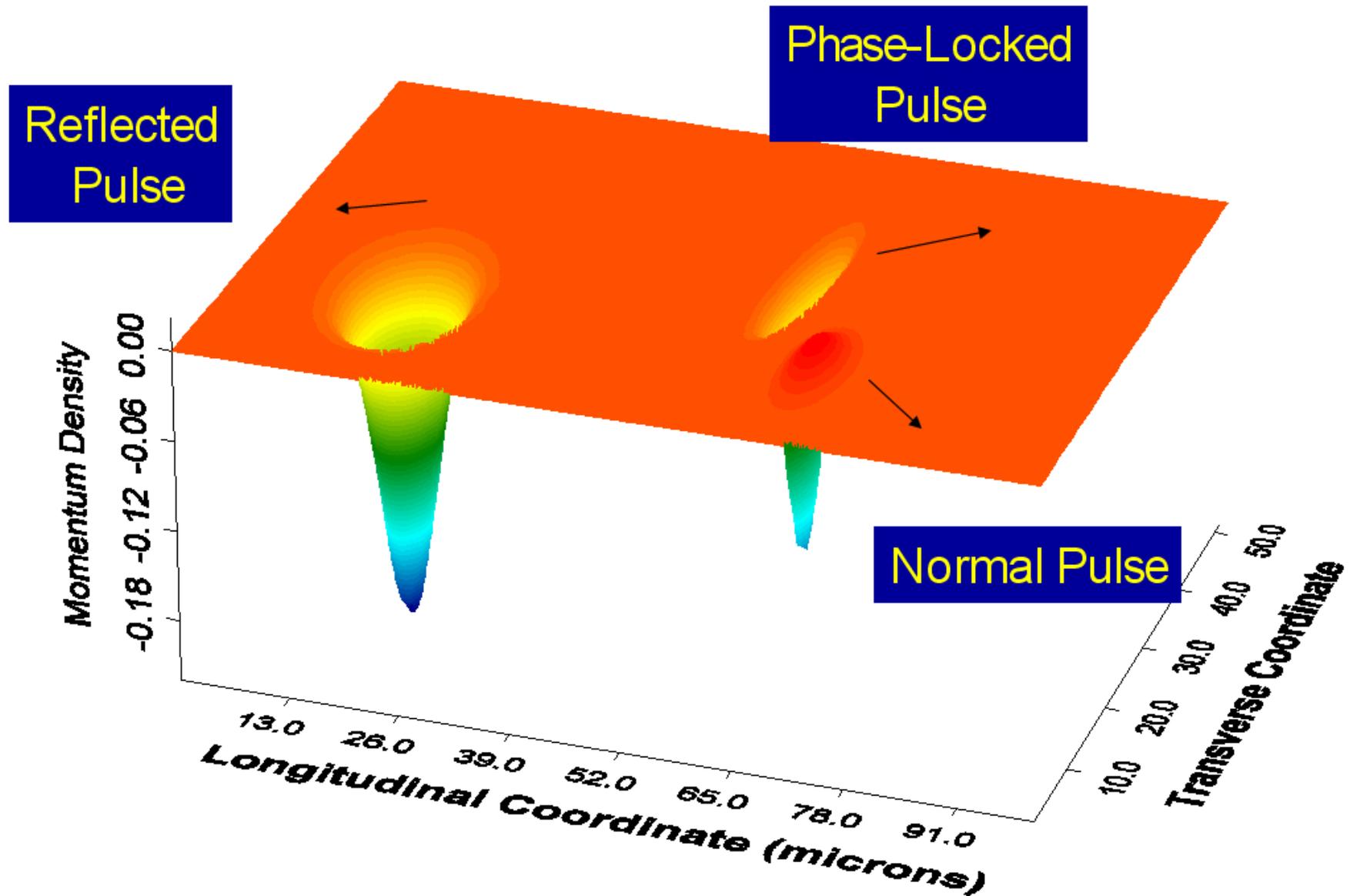

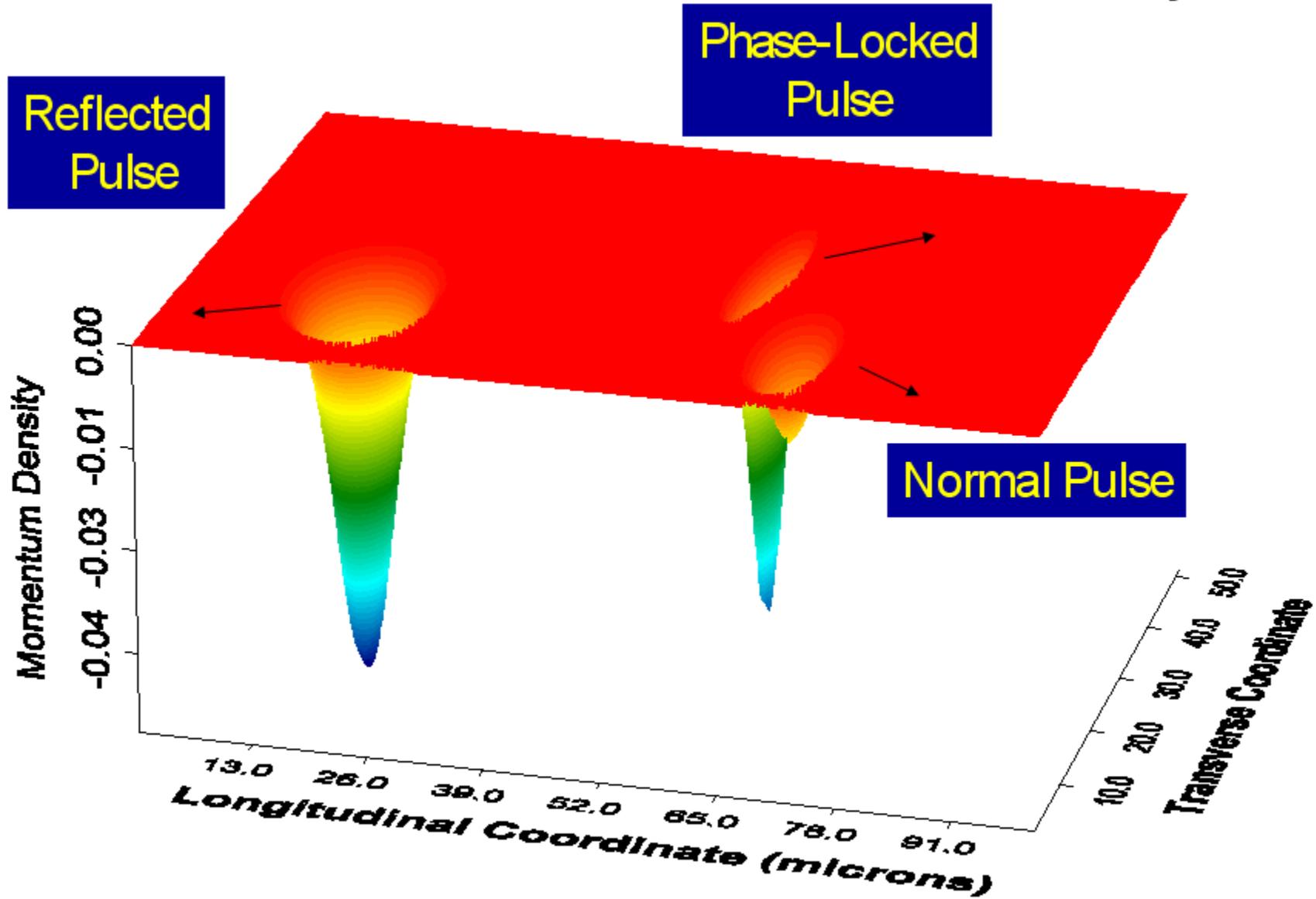